\documentclass[a4paper,11pt]{article}
\pdfoutput=1
\usepackage{jheppub} 
\usepackage[T1]{fontenc} 
\usepackage[english]{babel}
\usepackage[latin1]{inputenc}
\usepackage{amsmath,mathtools}
\usepackage{amsfonts}
\usepackage{amssymb}
\usepackage{wrapfig}
\usepackage{color}
\usepackage{float}
\usepackage{url}
\usepackage{mciteplus}
\usepackage{dsfont}
\usepackage{ifpdf}
\usepackage{epsfig}
\usepackage{xcolor}
\usepackage{hyperref}
\usepackage{cancel}
\usepackage{arydshln}
\usepackage{etoolbox}

\allowdisplaybreaks

\DeclareMathOperator{\Rgod}{\mathnormal{R}_\mathnormal{g}\o\check{\mathnormal{d}}}

\renewcommand{\a}{\alpha}
\renewcommand{\b}{\beta}
\newcommand{\g}{\gamma}
\newcommand{\G}{\Gamma}
\renewcommand{\k}{\kappa}

\renewcommand{\o}{\circ}
\newcommand{\Adg}{\text{Ad}_g}
\newcommand{\Adgi}{\text{Ad}_g^{-1}}

\newcommand{\Str}{\text{Str}}

\newcommand{\dP}{\check{d}}

\newcommand{\Rg}[1]{\mathnormal{R}_\mathnormal{g}\left(#1\right)}

\newcommand{\dds}{\mathnormal{d}^2\sigma\,}

\renewcommand{\d}{\delta}

\newcommand{\diag}{\text{diag}}

\newcommand{\algg}{\mathfrak{g}}
\newcommand{\algl}{\mathfrak{gl}}
\newcommand{\alggb}{\mathfrak{g}_{\text{b}}}
\newcommand{\alggf}{\mathfrak{g}_{\text{f}}}

\newcommand{\so}{\mathfrak{so}}
\newcommand{\su}{\mathfrak{su}}
\renewcommand{\u}{\mathfrak{u}}
\newcommand{\uosp}{\mathfrak{uosp}}
\newcommand{\psu}{\mathfrak{psu}}
\newcommand{\algi}[1]{\mathfrak{g}^{(#1)}}
\newcommand{\Ai}[1]{A^{(#1)}}

\newcommand{\Z}[1]{\mathbb{Z}_{#1}}

\newcommand{\Em}[1]{E^{#1}}

\newcommand{\Cmn}[2]{C_{#1}^{\;\;#2}}

\newcommand{\Lmn}[2]{\Lambda_{#1}^{\;\;#2}}

\newcommand{\Pb}[1]{P_2\left(#1\right)}
\newcommand{\jm}[1]{j^{#1}}

\newcommand{\Oi}{\mathcal{O}^{-1}}

\newcommand{\AdS}[1]{AdS_{#1}}
\newcommand{\Sph}[1]{S^{#1}}
\newcommand{\CP}[1]{\mathbb{CP}^{#1}}

\newcommand{\vphi}{\varphi}

\newcommand{\ELm}[1]{\textbf{L}_{#1}}
\newcommand{\EMa}[1]{\textbf{M}_{#1}}
\newcommand{\EL}{\textbf{L}}
\newcommand{\EK}[1]{\textbf{K}_{#1}}

\newcommand{\EH}{\textbf{H}}

\newcommand{\ET}[1]{\textbf{T}_{#1}}

\newcommand{\SOM}[1]{\textbf{M}_{#1}}

\newcommand{\ED}{\textbf{D}}

\newcommand{\el}[1]{\lambda_{#1}}

\newcommand{\es}[1]{\sigma_{#1}}

\renewcommand{\th}[1]{\theta_{#1}}
\newcommand{\dth}[1]{d\theta_{#1}}
\newcommand{\vf}[1]{\varphi_{#1}}

\newcommand{\dvf}[1]{d\varphi_{#1}}
\newcommand{\dps}{d\psi}
\newcommand{\dji}{d\xi}

\newcommand{\Et}[1]{\eta_{#1}}
\newcommand{\fabc}[3]{f_{#1#2}^{\quad#3}}
\newcommand{\eabc}[3]{\epsilon_{#1#2}^{\quad#3}}

\renewcommand{\G}[2]{G\!\left(#1\,|\,#2\right)}
\newcommand{\qs}[1]{q_{#1}^2}

\newcommand{\F}{\mathcal{F}}

\title{\boldmath 
Bosonic $\eta$-deformed $AdS_4\times\mathbb{CP}^3$ Background}

\author{Laura Rado,}
\author{Victor O. Rivelles}
\author{and Renato Sánchez}

\affiliation{Instituto de F\'{\i}sica, Universidade de S\~{a}o Paulo \\ Rua do Mat\~{a}o Travessa 1371, 05508-090 S\~{a}o Paulo, SP. Brazil}

\emailAdd{laura@if.usp.br}
\emailAdd{rivelles@fma.if.usp.br}
\emailAdd{renato@if.usp.br}

\abstract{
We build the bosonic $\eta$-deformed $\AdS{4}\times\CP{3}$ background generated by an $r$-matrix that satisfies the modified classical Yang-Baxter equation. In a special limit we find that it is the gravity dual of the noncommutative ABJM theory.

}

\begin{document}

\makeatletter
\patchcmd{\maketitle}{\@fpheader}{}{\hfill}{} 
\makeatother

\maketitle
\flushbottom

\section{Introduction}
A Hamiltonian system is integrable if there exists an infinite number of conserved charges in involution or, equivalently, if a Lax connection can be constructed. The required involution of these conserved charges leads to a particular form of the Poisson bracket of the Lax connection  
in terms of an $r$-matrix \cite{Maillet1985,Maillet1986,Maillet1986a}. In the context of string theory, the integrability of the $\AdS{5}\times\Sph{5}$ superstring  described by a $\sigma$-model on the supercoset ${\psu(2,2|4)}/{\so(1,4)\oplus\so(5)}$ \cite{Bena2004,Magro2009} is a remarkable property that also shows up in less symmetric cases like the $\AdS{4}\times\CP{3}$ background which is partially described by the supercoset ${\uosp(2,2|4)}/{\so(1,3)\oplus\u(3)}$ \cite{Arutyunov2008,Stefanski2009}. 

Since there is no generic and systematic way to construct integrable theories it is quite natural to look for deformations of known integrable theories which still preserve integrability. For the $\AdS{5}\times\Sph{5}$ superstring this has been extensively analyzed by adapting techniques used to deform integrable sigma models \cite{Delduc2013}.  
The strategy for building these deformations, called q-deformations, is to construct a Poisson bracket that preserves the relation between the Lax matrix and the undeformed Hamiltonian producing a deformed Hamiltonian while keeping the dependence of the Lax matrix on the currents.
From it we can derive a Lagrangian which is integrable and depends on the $R$-operator that satisfies the modified Classical Yang-Baxter equation (mCYBE). 
This procedure was applied to deform the $\AdS{5}\times\Sph{5}$ superstring \cite{Delduc2014} and it was found that its $\eta$-deformed background is not a solution of the standard type IIB supergravity equations despite the presence of $\kappa$-symmetry  \cite{Arutyunov2014,Arutyunov2015}.

For superstrings propagating in other backgrounds only a few results are known. Recently  some examples of integrable deformations of the $\AdS{4}\times\CP{3}$ background were given based on abelian solutions of the Classical Yang-Baxter equation (CYBE), which also have an interpretation in terms of TsT transformation \cite{Negron2018,Rado2020}. These deformed backgrounds are duals of noncommutative, dipole and $\beta$-deformed ABJM theory as well as a nonrelativistic limit having Schr\"odinger symmetry. 

In this paper we will consider deformations of the $\AdS{4}\times\CP{3}$ space based on a solution of the mCYBE whose $R$-operator is 
\begin{equation}
\label{REij}
R\left(E_{ij}\right)=\left\{\begin{matrix}
-iE_{ij} & \mathrm{if}& i<j\\ 
+iE_{ij}&  \mathrm{if}& i>j
\end{matrix}\right.,
\end{equation}
where $E_{ij}$, with $i,j=1,\dots,10$, are the $\algl(4|6)$ generators \cite{Kawaguchi2014}. Its main characteristic is that  the $r$-matrix, the map associated to the $R$-operator, 
is composed by the roots of the superalgebra. This is in contrast to the $r$-matrices that solve the CYBE which are given by the commuting generators of the superalgebra. 
Here we give the first steps by deriving the bosonic part of the deformed $\AdS{4}\times\CP{3}$ background via the standard coset construction based on the superalgebra $\algg=\uosp(2,2|6)$. 
We then take a special undeformed limit of this background that leads to the gravity dual of the noncommutative ABJM theory \cite{Rado2020}. 
Our results can be used towards to the computation of the full $\eta$-deformed $\AdS{4}\times\CP{3}$ background, which, once obtained, would allow us to explore a new family of integrable backgrounds, including those of a new type of generalized type IIA supergravity.

This paper is organized as follows. In \autoref{rmatrix} we present briefly the main steps needed to construct an $\eta$-deformed superstring $\sigma$-model. Next, in \autoref{cosetconstruction}, we review the coset construction of the undeformed bosonic $\AdS{4}\times\CP{3}$ background and in \autoref{news} we derive the $\eta$-deformed bosonic sector of ${\uosp(2,2|4)}/{\so(1,3)\oplus\u(3)}$. In \autoref{conclusions} we conclude and discuss our results.

\section{{$\eta$}-deformed Superstring Sigma Models \label{rmatrix}} 
The action for the $\eta$-deformed superstring $\sigma$-model on $\algg$ is  \cite{Delduc2014,Delduc2014a} 
\begin{equation}
\label{Sdef}
S=-\frac{\left(1+c\eta^2\right)^2}{4\left(1-c\eta^2\right)}\int \dds\left(\g^{\a\b}-\kappa\varepsilon^{\a\b}\right)\Str\left(A_\a,\dP J_\b\right),
\end{equation}
where $A=g^{-1}dg\;\in\algg$, $g\in G$, $\g^{\a\b}$ is the string worldsheet metric with $\det\g=1$  and $\k^2=1$. The $\Z{4}$-grading of $\algg$ allows the split of $A$ as
\begin{equation}
A=\Ai{0}+\Ai{1}+\Ai{2}+\Ai{3},\quad \left[\Ai{k},\Ai{m}\right]\subseteq\Ai{k+m}\:\text{mod}\:\Z{4}.
\end{equation}
The operator $\dP$ is defined 
as the following combination of projectors $P_i$ $(i=1,2,3)$ on the gradings of $\algg$
\begin{equation}
\label{d1}
\dP=P_1+\frac{2}{1-c\eta^2}P_2-P_3.
\end{equation}
The absence of $P_0$ is required for \eqref{Sdef} to be $\algi{0}$-invariant. The deformed current is then 
\begin{equation}
\label{J1}
J=\frac{1}{1-\eta\Rgod}A=\Oi A,
\end{equation}
where the operator $R_g$ is 
\begin{equation}
\label{Rgdef}
\Rg{M}=\Adgi\o R\o\Adg\left(M\right)=g^{-1}R(gMg^{-1})g,\qquad g\in G, 
\end{equation}
and $R\,:\algg\mapsto\algg$, which is the operator associated to the $r$-matrix required for integrability, must satisfy the Yang-Baxter equation (YBE) 
\begin{equation}\label{YBE}
\left[RM,RN\right]-R\left(\left[RM,N\right]+\left[M,RN\right]\right)=c\left[M,N\right],
\end{equation}
where $M,N\in\algg$. In \eqref{Sdef} and \eqref{YBE} the parameter $c$ refers to either  the classical Yang-Baxter equation (CYBE), $c=0$, or to the modified classical Yang-Baxter equation (mCYBE), $c=1$.  The case $c=1$, which is also known as non-split R-matrix \cite{Vicedo2015},  can be solved as \cite{Arutyunov2014},
\begin{equation}
\label{RMij}
R\left(M\right)_{ij}=-i\epsilon_{ij}M_{ij},\qquad \epsilon _{ij}=\left\{\begin{matrix}
1 & \mathrm{if} & i< j\\ 
 0&  \mathrm{if}& i=j\\ 
 -1& \mathrm{if}&i> j 
\end{matrix}\right.,\quad M\in\algg,
\end{equation}
and was considered in \cite{Delduc2014}. 
This type of deformation has been explored for superstrings in $\AdS{5}\times\Sph{5}$ \cite{Arutyunov2014,Arutyunov2015} but not for superstrings in $\AdS{4}\times\CP{3}$. In this work we will 
present some results concerning the bosonic $\eta$-deformed background based only on the bosonic roots of the algebra as done in \cite{Arutyunov2014}.

\section{Coset Construction of the Bosonic \texorpdfstring{$\AdS{4}\times\CP{3}$}{AdS4xCP3} Background \label{cosetconstruction}}


The isometry group of $\AdS{4}\times\CP{3}$ is the coset
\begin{equation}
\AdS{4}\times\CP{3}\equiv\frac{SO(2,3)}{SO(1,3)}\times\frac{SU(4)}{U(3)},
\end{equation}
which is part of the supercoset $UOSp(2,2|6)/\left(SO(1,3)\times U(3)\right)$ \cite{Arutyunov2008,Stefanski2009}. The supergroup $G=UOSp(2,2|6)$ has the superalgebra $\algg=\uosp(2,2|6)$ on which the $\sigma$-model can be constructed. The bosonic sector of $\algg$ can be expressed as \cite{Negron2018,Rado2020}
\begin{equation}
\label{algB}
\alggb:=\so(2,3)\oplus\su(2)\oplus\su(4)=\overbrace{\left(\so(1,3)\oplus\su(2)\oplus\u(3)\right)}^{\algi{0}}\oplus\overbrace{\left(\frac{\so(2,3)\oplus\su(2)\oplus\su{4}}{\so(1,3)\oplus\su(2)\oplus\u(3)}\right)}^{\algi{2}}.
\end{equation}
As supermatrices, the elements of $\algg$ can be written as
\begin{equation}
M_{(6|4)\times(6|4)}=\left(\begin{array}{c:c|c}
\so(2,3) & 0 & \overline{Q}\\ 
\hdashline
0 & \su(2)& 0\\ 
\hline
Q& 0 & \su(4)
\end{array}\right),
\end{equation}
where the dashed lines divide the supermatrix into the blocks corresponding to the subspaces $\AdS{4}$, $\CP{3}$ and $Q,\bar{Q}\in\alggf=\algi{1}\oplus\algi{3}$.

We will use the same basis as in \cite{Rado2020} to parametrize $\so(2,3)\oplus\su(2)\oplus\su(4)$ 
which, as supermatrices, can be expressed as
\begin{equation}
\label{supergens}
\SOM{ij}=\left(\begin{array}{c:c|c}
m_{ij} &  & \\ 
\hdashline
 & 0 & \\ 
\hline
 &  & 0
\end{array}\right),
\EMa{a}=-\frac{i}{2}\left(\begin{array}{c:c|c}
0 &  & \\ 
\hdashline
 & \es{a} & \\ 
\hline
 &  & 0
\end{array}\right),
\ELm{m}=-\frac{i}{2}\left(\begin{array}{c:c|c}
 0&  & \\ 
\hdashline
 & 0 & \\ 
\hline
 &  & \el{m}
\end{array}\right),
\end{equation}
where $m_{ij}$, with $i,j=0,1,\dots,4$, are the ten $4\times 4$ antisymmetric matrices representing the generators of isometries of $\so(2,4)$; $\es{a}$, with $a=1,2,3$ and $\el{m}$, with $m=1,\dots,15$, are, respectively, the usual $2\times 2$ Pauli and $4\times 4$ Gell-mann matrices of $\su(2)$ and $\su(4)$ (see \autoref{ApSO23} and \autoref{ApSU4}). 

The global isometries of the $\AdS{4}$ space can be written as 
\begin{equation}
\so(2,3)=\so(1,3)\oplus\frac{\so(2,3)}{\so(1,3)}, 
\end{equation}
where the coset $\frac{\so(2,3)}{\so(1,3)}$ is parametrized by
\begin{equation}
\label{EKAdS4}
\EK{0}=\frac{1}{2}\SOM{04},\quad\EK{1}=\frac{1}{2}\SOM{14},\quad\EK{2}=\frac{1}{2}\SOM{24},\quad\EK{3}=\frac{1}{2}\SOM{34} \equiv \frac{1}{2}\ED,
\end{equation}
where 
\begin{equation}
\Str\left(\EK{m}\EK{n}\right)=\frac{1}{4}\eta_{mn},\qquad m,n=0,1,2,3.
\end{equation}
An appropriate coset representative for $\AdS{4}$ is then 
\begin{equation}\label{gAdS4}
g_{\AdS{4}}=\exp\left(t\SOM{04}+\phi\SOM{12}\right)\exp\left(-\zeta\SOM{13}\right)\exp\left(\sinh^{-1}\rho\SOM{14}\right).
\end{equation}
The isometries of $\CP{3}$ space can be written as
\begin{equation}
\su(2)\oplus\su(4)=\su(2)\oplus\u(3)\oplus\frac{\su(2)\oplus\su(4)}{\su(2)\oplus\u(3)}.
\end{equation}
The coset $\frac{\su(2)\oplus\su(4)}{\su(2)\oplus\u(3)}$ can be described by 
\begin{equation}
\label{EKCP3}
\begin{gathered}
\EK{4}=\ELm{11},\quad\EK{5}=\ELm{12},\quad\EK{6}=\ELm{13},\\
\EK{7}=\ELm{14},\quad\EK{8}=\EH,\quad\EK{9}=\ELm{10},
\end{gathered}
\end{equation}
and 
$\EH=\ELm{6}+\ELm{9}+\EMa{1}$, with $\ELm{m}$ given in \eqref{supergens} and
\begin{equation}
\Str\left(\EK{m}\EK{n}\right)=\frac{1}{2}\d_{mn},\qquad m,n=4,\dots,9.
\end{equation}
Then an appropriate coset representative for $\CP{3}$ is \cite{Negron2018}
\begin{equation}\label{gCP3}
g_{\CP{3}}=\exp\left(\vphi_1\ELm{3}+\vphi_2\EL-\psi\EMa{3}\right)\exp\left(\theta_1\ELm{2}+(\theta_2+\pi)\ELm{14}\right)\exp\left(\left(2\xi+\pi\right)\left(\ELm{10}+\EMa{2}\right)\right),
\end{equation}
where 
\begin{equation}
\label{ELs}
\EL=-\frac{1}{\sqrt{3}}\ELm{8}-\sqrt{\frac{2}{3}}\ELm{15}.
\end{equation}
Hence, the bosonic coset representative for $\AdS{4}\times\CP{3}$ is
\begin{equation}
g_b=g_{\AdS{4}}\times g_{\CP{3}}.
\end{equation}

With this parametrization the undeformed $\AdS{4}\times\CP{3}$ metric\footnote{{We set $R_{\text{str}}^2=R^3/k=2^{5/2}\pi\sqrt{N/k}=1$, where $R_{\text{str}}^2$ is defined in \cite{Aharony2008}.}} has the expected form 
\begin{equation}
\label{AdS4metric}
ds^2_{\AdS{4}}=\frac{1}{4}\left(-\left(1+\rho^2\right)dt^2+\frac{d\rho^2}{1+\rho^2}+\rho^2\left(d\zeta^2+\cos^2\!\zeta d\phi^2\right)\right),
\end{equation}
and 
\begin{equation}
\label{CP3metric}
\begin{gathered}
ds^2_{\CP{3}}=\dji^2+\frac{1}{4}\cos^2\!\xi\left(\dth{1}^2+\sin^2\!\th{1}\dvf{1}^2\right)+\frac{1}{4}\sin^2\!\xi\left(\dth{2}^2+\sin^2\!\th{2}\dvf{2}^2\right)\\+\left(\frac{1}{2}\cos\th{1}\dvf{1}-\frac{1}{2}\cos\th{2}\dvf{2}+\dps\right)^2\sin^2\!\xi\cos^2\!\xi,
\end{gathered}
\end{equation}
where $\left(\th{1},\vf{1}\right)$ and $\left(\th{2},\vf{2}\right)$ parametrize the two spheres of $\CP{3}$, the angle $\xi$, $0\leq\xi\leq\pi/2$, determines their radii and $0\leq\psi\leq2\pi$ \cite{Cvetic2001}.

\section{The Bosonic \texorpdfstring{$\eta$}{eta}-deformed \texorpdfstring{$\AdS{4}\times\CP{3}$}{AdS4xCP3} Background\label{news}}

By switching off the fermionic degrees of freedom the deformed action in \eqref{Sdef} reduces to
\begin{equation}
\label{YBdef}
S=-\frac{1}{2}\left(\frac{1+\eta^2}{1-\eta^2}\right)^2\int \dds\left(\g^{\a\b}-\varepsilon^{\a\b}\right)\Str\left(A_\a,\Pb{J_\b}\right),\quad\k=1.
\end{equation}
The action of $P_2$ on $A$, $\Rg{\EK{m}}$ and $J$ are \cite{Negron2018,Rado2020}
\begin{equation}
\label{P2AJ}
\Pb{A}=\Em{m}\EK{m},\quad\Pb{\Rg{\EK{m}}}=\Lmn{m}{n}\EK{n},\quad\Pb{J}=\jm{m}\EK{m},
\end{equation}
where $\EK{m}$ are the generators of $\algi{2}$ and the coefficients $\jm{m}$ can be obtained from
\begin{equation}
\label{jms}
\jm{m}=\EK{n}\Cmn{n}{m}. 
\end{equation}
The coefficients $\Cmn{n}{m}$ can be seen as
a matrix in terms of $\Lmn{m}{n}$
\begin{equation}
\mathbf{C}=\left(\mathbf{I}-\chi\mathbf{\Lambda}\right)^{-1},\quad\chi=\frac{2\eta}{1-\eta^2}.
\end{equation}
Then, from \eqref{YBdef}, we can read off the metric and the B-field as 
\begin{equation}\label{Gmn}
ds^2=\Str\left(A\,\Pb{J}\right)=\frac{1}{4}\sum_{m=0}^{3}\jm{m}\Str\left(A\EK{m}\right)+\frac{1}{2}\sum_{m=4}^{9}\jm{m}\Str\left(A\EK{m}\right),
\end{equation} 
\begin{equation}\label{Bmn}
B=\Str\left(A\wedge\Pb{J}\right)=-\frac{1}{4}\sum_{m=0}^{3}\jm{m}\wedge\Str\left(A\EK{m}\right)-\frac{1}{2}\sum_{m=4}^{9}\jm{m}\wedge\Str\left(A\EK{m}\right).
\end{equation}

{
Now, in order to compute the deformed background with the action \eqref{YBdef} we have first to compute the the nonzero components of $\Lmn{m}{n}$ in \eqref{P2AJ}
\begin{equation}
\label{lambdas}
\begin{gathered}
\Lmn{0}{1}=\Lmn{1}{0}=-\rho,\\
\Lmn{2}{3}=-\Lmn{3}{2}=-\rho\sin\zeta,\\
\Lmn{4}{5}=-\Lmn{5}{4}=\sin^2\left(\frac{\th{1}}{2}\right)-\cos2\xi\cos^2\left(\frac{\theta_1}{2}\right)=q_1,\\
\Lmn{5}{8}=\Lmn{8}{5}=-\sin\th{1}\sin\xi\cos^2\xi=q_2,\\
\Lmn{6}{7}=-\Lmn{7}{6}=-\cos^2\left(\frac{\th{2}}{2}\right)-\cos2\xi\sin^2\left(\frac{\th{2}}{2}\right)=q_3,\\
\Lmn{7}{9}=-\Lmn{8}{7}=-\sin\th{2}\cos\xi\sin^2\xi=q_4,\\
\Lmn{8}{9}=-\Lmn{9}{8}=-\cos2\xi=q_5.
\end{gathered}
\end{equation}
Then we follow the same procedure as in \cite{Rado2020} to find the $\eta$-deformed metric. The $\AdS{4}$ part is given by  
\begin{equation}
\label{AdS4def}
\frac{ds^2_{\AdS{4}}}{1+\chi^2}=\frac{1}{4}\left(-\frac{f_+(\rho)}{f_-(\chi\rho)}dt^2+\frac{1}{f_+(\rho)f_-(\chi\rho)}d\rho^2+\frac{1}{1+\chi^2\rho^4\sin^2\zeta}\rho^2 ds^2_{\Sph{2}}\right),
\end{equation}
where $f_\pm(x)=1\pm x^2$ and $ds^2_{\Sph{2}}$ is an undeformed $\Sph{2}$ sphere parametrized by $(\zeta,\phi)$. For the $\CP{3}$ part we find 
\begin{equation}
\label{CP3def}
\begin{gathered}
\frac{ds^2_{\CP{3}}}{1+\chi^2}=\F\bigg[\G{\qs{1}+\qs{2}+\qs{3}+\qs{4}}{\qs{1}\qs{3}+\qs{1}\qs{4}+\qs{2}\qs{3}}\dji^2+\G{\qs{3}+\qs{4}+\qs{5}}{\qs{3}\qs{5}}\frac{1}{4}\cos^2\!\xi\dth{1}^2\\
+\G{\qs{2}+\qs{3}+\qs{4}+\qs{5}}{\qs{2}\qs{3}+\qs{3}\qs{5}}\frac{1}{4}\cos^2\!\xi\sin^2\!\th{1}\dvf{1}^2+\G{\qs{1}+\qs{2}+\qs{5}}{\qs{1}\qs{5}}\frac{1}{4}\sin^2\!\xi\dth{2}^2\\
+\G{\qs{1}+\qs{2}+\qs{4}+\qs{5}}{\qs{1}\qs{4}+\qs{1}\qs{5}}\frac{1}{4}\sin^2\!\xi\sin^2\!\th{2}\dvf{2}^2\\
+\G{\qs{1}}{0}\G{\qs{3}}{0}\left(\frac{1}{2}\cos\th{1}\dvf{1}-\frac{1}{2}\cos\th{2}\dvf{2}+\dps\right)^2\sin^2\!\xi\cos^2\!\xi\bigg],
\end{gathered}
\end{equation}
where $q_1, \dots q_5$ are given in \eqref{lambdas},
\begin{equation}
\F^{-1}=\G{\qs{1}+\qs{2}+\qs{3}+\qs{4}+\qs{5}}{\qs{1}\qs{3}+\qs{1}\qs{4}+\qs{1}\qs{5}+\qs{2}\qs{3}+\qs{3}\qs{5}},
\end{equation}
and 
\begin{equation}
\G{r}{s}=1+r\chi^2+s\chi^4.
\end{equation}
Notice that for the undeformed theory, when $\chi=0$, we get $\F=\G{r}{s}=1$ 
and the metrics \eqref{AdS4def} and \eqref{CP3def} reduce to \eqref{AdS4metric} and \eqref{CP3metric} respectively.
Finally, the bosonic part of the $\eta$-deformed $B$-field is given by 
\begin{equation}
\label{Bdef}
\begin{gathered}
\frac{B}{1+\chi^2}=\frac{1}{2}\left(\frac{\chi\rho}{1-\chi^2\rho^2}dt\wedge d\rho\right)+\frac{1}{2}\left(\frac{\chi\rho^4\cos\zeta}{1+\chi^2\rho^4\sin^2\!\zeta}\right)d\zeta\wedge d\phi\\
+\F\bigg(\frac{1}{2}q_2q_5\chi^2\G{\qs{3}}{0}\cos\xi d\xi\wedge\dth{1}+\frac{1}{2}q_4q_5\chi^2\G{\qs{3}}{0}\sin\xi\dji\wedge\dth{2}\\
+q_5\chi\G{\qs{3}}{0}\cos\xi\left(\G{\qs{1}}{0}\cos\th{1}\sin\xi+q_1q_2\chi^2\sin\th{1}\right)\dji\wedge\dvf{1}\\
-q_5\chi\G{\qs{1}}{0}\sin\xi\left(\G{\qs{3}}{0}\cos\th{2}\cos\xi+q_3q_4\chi^2\sin\th{2}\right)\dji\wedge\dvf{2}\\
+q_5\chi\G{\qs{1}}{0}\G{\qs{3}}{0}\sin2\xi\dji\wedge\dps\\
+\frac{\chi}{2}\left(q_1\G{\qs{3}+\qs{4}+\qs{5}}{\qs{3}\qs{5}}\sin\th{1}-q_2\G{\qs{3}}{0}\cos\th{1}\sin\xi\right)\cos^2\!\xi\dth{1}\wedge\dvf{1}\\
+\frac{\chi}{4}\left(q_3q_4\chi^2\sin\th{2}+\G{\qs{3}}{0}\cos\th{2}\cos\xi\right)\sin2\xi\dth{1}\wedge\dvf{2}\\
-q_2\chi\G{\qs{3}}{0}\sin\xi\cos^2\!\xi\dth{1}\wedge\dps\\
+\frac{\chi}{4}\left(q_1q_2\chi^2\sin\th{1}+\G{\qs{1}}{0}\cos\th{1}\sin\xi\right)\sin2\xi\dth{2}\wedge\dvf{1}\\
+\frac{\chi}{2}\left(q_3\G{\qs{1}+\qs{2}+\qs{5}}{\qs{1}\qs{5}}\sin\th{2}-q_4\G{\qs{1}}{0}\cos\th{2}\cos\xi\right)\sin^2\!\xi\dth{2}\wedge\dvf{2}\\
+q_4\chi\G{\qs{1}}{0}\sin^2\!\xi\cos\xi\dth{2}\wedge\dps
\bigg),
\end{gathered}
\end{equation}
which vanishes for 
the undeformed theory. 

The deformed background \eqref{AdS4def}, \eqref{CP3def} and \eqref{Bdef} breaks the isometries of the $AdS$ space so that the dual gauge theory will be neither conformal nor Lorentz invariant. Also, the presence of the $B$-field indicates that the dual gauge theory would be a noncommutative deformation of the ABJM theory.

We can now find a special undeformed limit which is a solution of the standard supergravity in a similar way as has been done for $\AdS{5}\times\Sph{5}$ \cite{Arutyunov2015}.  We first rescale the $\AdS{4}$ coordinates 
\begin{equation}
t\rightarrow\sqrt{\chi}t,\;\rho\rightarrow\rho/\sqrt{\chi},\;\zeta\rightarrow\zeta_0+\sqrt{\chi}\zeta,\;\phi\rightarrow\sqrt{\chi}\phi/\cos\zeta_0,
\end{equation}
where $\zeta_0$ is a parameter, and then set $\chi=0$ to get 
\begin{equation}
ds^2=\frac{1}{4}\left(-\rho^2dt^2+\frac{d\rho^2}{\rho^2}+\frac{\rho^2}{1+\rho^4\sin^2\zeta_0}\left(d\zeta^2+d\phi^2\right)\right)+ds^2_{\CP{3}}, 
\end{equation}
and 
\begin{equation}
B=\frac{1}{2}\frac{\rho^4\sin\zeta_0}{1+\rho^4\sin^2\zeta_0}d\zeta\wedge d\phi.
\end{equation}
This is the gravity dual of the noncommutative ABJM theory with deformation parameter $\sin\zeta_0$  \cite{Rado2020} which is a type IIA supergravity solution. The connection between the $\eta$-deformed background and the gravity dual of the noncommutative ABJM theory can be understood as an infinite boost of the $r$-matrix that generates the $\eta$-deformed background which gives a $r$-matrix with parameter $\sin\zeta_0$ leading to the gravity dual of the noncommutative theory \cite{Hoare2016b}.


\section{Conclusions and Outlook \label{conclusions}}

In this paper we derived the bosonic $\eta$-deformed $\AdS{4}\times\CP{3}$ background using the same technique developed for  the bosonic Yang-Baxter deformed $\AdS{4}\times\CP{3}$ background \cite{Rado2020}. We have also shown that in the special limit $\chi\rightarrow0$  it is the the gravity dual of the noncommutative ABJM theory. In the $\AdS{5}\times S^5$ case there is also the so-called ``maximal deformation'' limit $\chi\rightarrow\infty$ \cite{Arutyunov2014a} that in our case should leads to the mirror $\AdS{4}\times\CP{3}$ background. This would allow us to study the finite size thermodynamic Bethe ansatz. However, in our case, the construction of the double Wick rotated background that generates the mirror undeformed $\AdS{4}\times\CP{3}$ background is not simply the interchange of the metric elements corresponding to the coordinates involved in the light-cone gauge fixing \cite{Klose2012}. To do that we have to take coordinate $t$ in $\AdS{4}$ and the  coordinate $\psi$ in $\CP{3}$.  This happens because $\CP{3}$ is not block diagonal so that the double Wick rotation constructed in \cite{Arutyunov2014a} (see also \cite{Arutyunov2014b,Arutyunov2015a}) must be adapted to our case. After that, it will be possible to get the maximal deformation of our $\eta$-deformed metric and $B$-field and see if it also has the mirror background as a limit. Another interesting limit is $\chi\rightarrow i$, which establishes the connection to the Pohlmeyer reduced model for the undeformed $\AdS{4}\times\CP{3}$ \cite{Hoare2014}. These are important topics that deserve further study.

The deformed fermionic sector can in principle be obtained in the same way as in the $\AdS{5}\times\Sph{5}$ case \cite{Arutyunov2015} but there are important subtleties that must be taken into account. The superalgebra $\uosp(2,2|6)$ does not describe all fermionic degrees of freedom of the $\AdS{4}\times\CP{3}$ string \cite{Arutyunov2008} which are needed to build the RR sector. A fermionic factor $g_f$ will appear in the coset representatives needed to obtain the fermionic currents $A^{(1)}$, $A^{(3)}$  and the fermionic contributions to the operator $\Oi$ in \eqref{J1}. Then they can be expanded up to quadratic order to get the deformed fluxes of the theory.
The first problem one must face to get the full deformed background is to reincorporate those 32 fermions. Since not all of them are part of the supercoset, a possibility is to start from a supercoset that contains them. 

The $R$-operator \eqref{RMij} was also considered in $\AdS{5}\times\Sph{5}$ \cite{Arutyunov2014} where it was found that the deformed background does not solve the standard type IIB supergravity equations but a kind of generalized supergravity equations \cite{Arutyunov2016,Wulff2016}. We then expect that our NSNS fields \eqref{AdS4def}, \eqref{CP3def} and \eqref{Bdef} are also part of some  generalized supergravity. Besides that, the $R$-operator \eqref{RMij}, called reference $R$-operator in \cite{Hoare2019}, can be associated to a $r$-matrix which is not unimodular \cite{Borsato2016a}. However, as shown in \cite{Hoare2019}, it is possible to find a permutation of the $R$-operator which gives an unimodular inequivalent $R$-operator such that the $\eta$-deformed background is a solution of type IIB supergravity with the same NSNS fields. It will be very interesting to find out whether the $\AdS{4}\times\CP{3}$ case has the same property.


\appendix

\section{A Basis for the \texorpdfstring{$\so(2,3)$}{so(2,3)} Algebra \label{ApSO23}} 
The 10 generators of $SO(2,3)$ can be written as
\begin{equation}\label{gens}
m_{ij}=\frac{i}{4}\left[\Gamma_i,\Gamma_j\right],
\end{equation}
and satisfy
\begin{equation}
\left[m_{ij},m_{k\ell}\right]=i\left(\eta_{i\ell}m_{jk}+\eta_{jk}m_{i\ell}-\eta_{j\ell}m_{ik}-\eta_{ik}m_{j\ell}\right),
\end{equation}
where $i,j,k,\ell=0,1,2,3,4$. We choose the following representation for the $SO(2,3)$ $\Gamma_i$ matrices
\begin{equation}
\left\{\Gamma_i,\Gamma_j\right\}=2\eta_{ij}, 
\end{equation}
\begin{equation}
\Gamma_i=\left\{\begin{matrix}
i\gamma_5\gamma_a & i=a=0,1,2,3\\ 
\gamma_5=i\gamma_0\gamma_1\gamma_2\gamma_3 & i=4
\end{matrix}\right.
\end{equation}
with $\eta_{ij}=\text{diag}(-+++-)$, and $\gamma_a$ being the gamma matrices in a Dirac representation of $SO(1,3)$  \cite{Fabbri2000} (see \cite{Arutyunov2008} for a different choice)
\begin{equation}
\begin{gathered}
\gamma_0=\begin{pmatrix}
 I_2 &  0  \\ 
 0   & -I_2 \\  
\end{pmatrix},\quad
\gamma_1=\begin{pmatrix}
 0 &  \sigma_3  \\ 
 -\sigma_3   & 0 \\  
\end{pmatrix},\\
\gamma_2=\begin{pmatrix}
 0 &  \sigma_1  \\ 
 -\sigma_1   & 0 \\  
\end{pmatrix},\quad
\gamma_3=\begin{pmatrix}
 0 &  \sigma_2  \\ 
 -\sigma_2 & 0 \\  
\end{pmatrix}.
\end{gathered}
\end{equation}
and 
\begin{equation}
\gamma_5=\begin{pmatrix}
 0 &  I_2  \\ 
I_2 & 0 \\  
\end{pmatrix}.
\end{equation}
From \eqref{gens}, we get
\begin{equation}
m_{ab}=\frac{1}{4}\left[\gamma_a,\gamma_b\right],\qquad  m_{a4}=\frac{i}{2}\gamma_a,\qquad a,b=0,1,2,3.
\end{equation}

In order to make explicit the conformal group let us split the indices as
\begin{eqnarray}
m_{ij}=\left\{m_{\mu\nu},m_{\mu3},m_{\mu4},m_{34}\right\},\qquad  \mu,\nu=0,1,2,
\end{eqnarray}
such that $\eta_{\mu\nu}=\text{diag}(-,+,+)$ \footnote{This is going to be the signature on the Minkowskian boundary of $\AdS{4}$.}. Let us also define \cite{Fabbri2000}
\begin{align}\nonumber
p_\mu&=m_{\mu4}+m_{\mu3}, \\ \nonumber
k_\mu&=m_{\mu4}-m_{\mu3}, \\
D&=m_{34}.
\end{align}
Then the conformal algebra $SO(2,3)$ is
\begin{align}\nonumber
\left[m_{\mu\nu},m_{\rho\sigma}\right]&=\eta_{\mu\sigma}m_{\nu\rho}+\eta_{\nu\rho}m_{\mu\sigma}-\eta_{\mu\rho}m_{\nu\sigma}-\eta_{\nu\sigma}m_{\mu\rho},\\
\nonumber
\left[m_{\mu\nu},D\right]&=0,\\
\nonumber
\left[D,p_\mu\right]&=-p_\mu,\\
\label{conformal4}
\left[D,k_\mu\right]&=k_\mu,\\
\nonumber
\left[k_\mu,p_\nu\right]&=2\eta_{\mu\nu}D+2m_{\mu\nu},\\
\nonumber
\left[m_{\mu\nu},p_\rho\right]&=-\eta_{\mu\rho}p_\nu+\eta_{\nu\rho}p_\mu,\\
\nonumber
\left[m_{\mu\nu},k_\rho\right]&=-\eta_{\mu\rho}k_\nu+\eta_{\nu\rho}k_\mu.
\end{align}

\section{A Basis for the \texorpdfstring{$\su(4)$}{su(4)} Algebra \label{ApSU4}}

A basis for $\su(4)$ can be constructed in terms of anti-hermitian $4\times 4$ matrices known as Gell-Mann matrices,
\begin{align}\nonumber
\lambda_1=& \begin{pmatrix}
0 &1  &0  &0 \\ 
1 &0  &0 & 0\\ 
 0& 0 &  0&0 \\ 
 0&0  &0  &0 
\end{pmatrix},   \qquad  \lambda_2=\begin{pmatrix}
0 &-i  &0  &0 \\ 
i &0  &0 & 0\\ 
 0& 0 &  0&0 \\ 
 0&0  &0  &0 
\end{pmatrix}, \qquad \lambda_3=\begin{pmatrix}
1 &0  &0  &0 \\ 
0 &-1  &0 & 0\\ 
 0& 0 &  0&0 \\ 
 0&0  &0  &0 
\end{pmatrix},
\\
\nonumber
\lambda_4=& \begin{pmatrix}
0 &0  &1  &0 \\ 
0 &0  &0 & 0\\ 
 1& 0 &  0&0 \\ 
 0&0  &0  &0 
\end{pmatrix},   \qquad  \lambda_5=\begin{pmatrix}
0 &0  &-i  &0 \\ 
0 &0  &0 & 0\\ 
 i& 0 &  0&0 \\ 
 0&0  &0  &0 
\end{pmatrix}, \qquad \lambda_6=\begin{pmatrix}
0 &0  &0  &0 \\ 
0 &0  &1 & 0\\ 
 0& 1 &  0&0 \\ 
 0&0  &0  &0 
\end{pmatrix},
\\ 
\nonumber
\lambda_7=& \begin{pmatrix}
0 &0  &0  &0 \\ 
0 &0  &-i & 0\\ 
 0& i &  0&0 \\ 
 0&0  &0  &0 
\end{pmatrix},   \qquad  \lambda_8=\frac{1}{\sqrt{3}}\begin{pmatrix}
1 &0  &0  &0 \\ 
0 &1  &0 & 0\\ 
 0& 0 &  -2&0 \\ 
 0&0  &0  &0 
\end{pmatrix}, \qquad \lambda_9=\begin{pmatrix}
0 &0  &0  &1 \\ 
0 &0  &0 & 0\\ 
 0& 0 &  0&0 \\ 
 1&0  &0  &0 
\end{pmatrix},
\\
\nonumber
\lambda_{10}=& \begin{pmatrix}
0 &0  &0  &-i \\ 
0 &0  &0 & 0\\ 
 0& 0 &  0&0 \\ 
 i&0  &0  &0 
\end{pmatrix},   \qquad  \lambda_{11}=\begin{pmatrix}
0 &0  &0  &0 \\ 
0 &0  &0 & 1\\ 
 0& 0 &  0&0 \\ 
 0&1  &0  &0 
\end{pmatrix}, \qquad \lambda_{12}=\begin{pmatrix}
0 &0  &0  &0 \\ 
0 &0  &0 & -i\\ 
 0& 0 &  0&0 \\ 
 0&i  &0  &0 
\end{pmatrix},
\\
\lambda_{13}=& \begin{pmatrix}
0 &0  &0  &0 \\ 
0 &0  &0 & 0\\ 
 0& 0 &  0&1 \\ 
 0&0  &1 &0 
\end{pmatrix},   \qquad  \lambda_{14}=\begin{pmatrix}
0 &0  &0  &0 \\ 
0 &0  &0 & 0\\ 
 0& 0 &  0&-i \\ 
 0&0  &i  &0 
\end{pmatrix}, \qquad \lambda_{15}=\frac{1}{\sqrt{6}}\begin{pmatrix}
1 &0  &0  &0 \\ 
0 &1  &0 & 0\\ 
 0& 0 &  1&0 \\ 
 0&0  &0  &-3 
\end{pmatrix}.
\end{align}
The first $8$ matrices form a basis for $\su(3)\subset \su(4)$. Furthermore, these matrices are orthogonal and satisfy
\begin{equation}
\mathrm{Tr} \left(\lambda_m \lambda_n\right)=2\delta_{mn},\qquad m=1,...,15,
\end{equation}
and commutation relations
\begin{equation}
\left[\lambda_m,\lambda_n\right]=2i f_{mn}^{p}\lambda_p.
\end{equation}
A list of non-vanishing structure constants can be found in \cite{Pfeifer2003}. In this representation the Cartan generators are given by $\lambda_3$, $\lambda_8$ and $\lambda_{15}$.

\acknowledgments
 We wish to thank Stijn van Tongeren for comments on the deformation limits. The work of V. O. Rivelles was supported by FAPESP grant 2019/21281-4.

\bibliographystyle{JHEP}
\bibliography{Biblioteca}
\end{document}